\documentclass[12pt]{article}
\usepackage{latexsym}
\usepackage{mathcomp}
\usepackage{graphicx}
\usepackage{graphics}
\usepackage{epstopdf}
\usepackage{epsfig}

\usepackage{graphicx}
\usepackage{float}
 \usepackage{amsmath}
\usepackage{amsthm}
\usepackage{amssymb}
\usepackage{amscd}
\usepackage{amsfonts}
\usepackage{makeidx}

 \newcommand{\beqn}{\begin{eqnarray}}
 \newcommand{\eeqn}{\end{eqnarray}}
 \newcommand{\be}{\begin{equation}}
 \newcommand{\ee}{\end{equation}}
 \newcommand{\ba}{\begin{array}}
 \newcommand{\ea}{\end{array}}

 \newcommand{\re}{\ref}
 \newcommand{\ci}{\cite}
 \newcommand{\la}{\label}
 \newcommand{\bfr}{\begin{flushright}}
 \newcommand{\efr}{\end{flushright}}
 \newcommand{\bfl}{\begin{flushleft}}

 \newcommand{\efl}{\end{flushleft}}
 \newcommand{\fr}{\frac}
 \newcommand{\ov}{\overline}
 
 \newcommand{\vp}{\varphi}
 \newcommand{\ve}{\varepsilon}
 \newcommand{\de}{\delta}

 \newcommand{\al}{\alpha}
 
 \newcommand{\om}{\omega}

 \newcommand{\ds}{\displaystyle}

\newcommand{\no}{\noindent}
 \newcommand{\br}{|\kern-.25em|\kern-.25em|}
 \newcommand{\brr}{{|\kern-.15em|\kern-.15em|\kern-.15em}\,}
 \newcommand{\ddd}{\st{.\kern-.07em.\kern-.07em.}}

\newcommand{\bo}{{\hfill\loota}}
\newcommand{\loota}{\hbox{\enspace{\vrule height 7pt depth 0pt width
      7pt}}}

 \def\N{{\rm I\kern-.1567em N}}                              

\def\R{{\rm I\kern-.1567em R}}                              
 \def\C{{\rm C\kern-4.7pt                                    
 \vrule height 7.7pt width 0.4pt depth -0.5pt \phantom {.}}}
 \def\Z{{\sf Z\kern-4.5pt Z}}                                
 
 \def\Im {{\rm Im\,}}                                        

 \newcommand{\supp}{\mathop{\rm supp}\nolimits}

 \newtheorem{theorem}{Theorem}[section]
 
 \newtheorem{definition}[theorem]{Definition}
 
 \newtheorem{lemma}[theorem]{Lemma}
 \newtheorem{example}[theorem]{Example}
 \newtheorem{remark}[theorem]{Remark}
 
 \newtheorem{corollary}[theorem]{Corollary}
 \newtheorem{proposition}[theorem]{Proposition}

\begin{document}

 \vspace{2cm}

\vspace{3cm}
\title
{\huge\bf On justification of
Sobolev's formula
\medskip
for diffraction  by wedge}
\vspace{1cm}

\maketitle

\author{
{\Large A.I.Komech \footnote{
Supported partly by Alexander von Humboldt Research Award, Austrian
Science Fund (FWF): P22198-N13, and RFBR.
}}\\
{\it
Faculty of Mathematics of Vienna University\\
and the Institute for the Information Transmission Problems of
RAS (Moscow)}\\
e-mail:~alexander.komech@univie.ac.at
\medskip\\
{\large A.E.~Merzon and  \large J.E. De la Paz ~M\'{e}ndez
\footnote{Supported  by CONACYT and CIC of
 UMSNH and FWF-project P22198-N13.}}\\
{\it Institute of Physics and  Mathematics\\
University of Michoac\'{a}n of
 San Nicolas de Hidalgo \\
 Morelia, Michoac\'{a}n, M\'{e}xico }\\
  e-mail:~anatoli@ifm.umich.mx.
}
 \pagenumbering{arabic}
\setcounter{page}{1}

 \section{Introduction}

\no In this paper we
develop our results \cite {md3}-\cite {eam}  on scattering of plane waves by two-dimensional wedge
\be\la{W}
W:=\{
y=(y_1,y_2)\in\R^2: y_1=\rho\cos\theta,~~y_2=\rho\sin\theta,~\rho\ge 0,
~0\le \theta\le\phi\}
\ee
 of the magnitude $\phi\in (0,\pi)$.
In
these papers the scattering
 was studied for the harmonic incident  waves
\be\la{uin}
 u_{\rm in}(y,t)=e^{-i\omega_0 (t- n_0\cdot y)} f(t-n_0\cdot y){\rm
 ~~~~~for~~}t\in\R {\rm ~~and~~} y\in Q,
\ee
where
$n_0=(\cos\alpha,\sin\alpha)$  and
$Q:=\R^2\setminus W$
is the angle of the magnitude
\begin{equation}\label{sp}
\Phi:=2\pi-\phi,~~~\Phi\in(\pi,2\pi).
\end{equation}
The boundary  $\partial Q=Q_1\cup Q_2\cup 0$, where
\be\la{ass1}
Q_1:=\{(y_1,0):y_1>0\} ~~{\rm and}~~
Q_2:=\{(\rho\cos\phi,\rho\sin\phi):\rho>0\}.
\ee
Further,
the profile function $f$ is a Heaviside-type smooth function:
\be\la{f}
f\in C^{\infty}(\R), ~~~{\rm supp}~ f\subset [0,\infty),~~~ \mbox{and}~~~~
f(s)=1~ \mbox{for}~~s\geq s_0
\ee
where $s_0>0$. The diffraction is described by the mixed
problem
\begin{equation}\label{sidlu}
\left\{
\ba{l}
\Box u(y,t)=0,~y\in Q;~~~B u(y,t)\vert_{Q_1\cup Q_2}=0,~t\in \R\\
u(y,t)=u_{\rm in}(y,t),~y\in Q,~t<0.
\ea
\right .
\end{equation}
Here $\Box=\partial_{t}^2-\bigtriangleup$, $B=(B_1,B_2)$, and $B u\vert_{Q_1\cup Q_2}=(B_1 u\vert_{Q_1},B_2 u\vert_{Q_2})$,
where $B_{1,2}$ are equal to either the identity operator $I$ or to $\partial/\partial n$
 where $n$ is the outward normal to $Q$.
The $DD$-problem corresponds to
 $B_1=B_2=I$,
 the $NN$-problem corresponds to
 $B_1=B_2=\partial/\partial n,
 $
 and the $DN$-problem corresponds to
 $B_2=I, B_1=\partial/\partial n.$

The uniqueness, existence, the formula for solution  to
(\ref{sidlu}) and the Limiting Amplitude Principle were proved
in \cite{md3}-\cite{eam}.
Now we generalize these results to the case of nonsmooth and nonperiodic incident wave
\begin{equation}\label{dia}
u_{\rm in}(y,t)=F(t-n_0\cdot y),~~~y\in\R^2, ~~~t\in\R,
\end{equation}
where $F$ is a tempered distribution
with support in ${\ov {\R^+}}$.

Our main results are the formulas for
solutions to the nonstationary
problems (\re{sidlu})
\be\label{svertka-i}
u=u_{\rm in}+ F_\de*{\cal J},~~
(y,t)\in \overline Q\times  \R,
\ee
where ${\cal J}$ is a suitable distribution corresponding
to the type of boundary conditions either $DD$, or $NN$, or $DN$.
Here
$F_\de(y,t):=F(t)\de (y)$, and the convolution is well defined
in the sense of distributions (see Theorem \ref{sush} for the  $DD$ case).

As an application, we reproduce
the Sobolev formula
obtained in \cite {Sob}
for $F(s)=h(s)$
(Heaviside function) in the case of the DD-problem.
We also obtain similar formulas for the NN- and DN problems.

Moreover, we give the explicit formula for solution for $F(s)=\delta(s)$.
We  study also the case when
 $F$ is a locally summable function such that
\begin{equation}\label{F}
F(s)=0,~~s<0,~~\sup (1+|s|)^p|F(s)|<\infty,~~s\in\R
\end{equation}
for some $p\in \R$. We analyze the stabilization
of solutions as $t\to \infty$.
Namely, we prove that the solution
locally tends to a limit
as $t\to\infty$ if and only if
$F(s)\to C$, $s\to \infty$.

We also generalize the Limiting Amplitude Principle
which was proved in \cite{kmm} for
 smooth Heaviside-type incident waves:
in  (\ref{uin}) for the DD case,
and in \cite{MM1,mm} and \cite {eam} for the DN- and NN cases respectively.
Namely, we consider the incident waves with
$F(s)-a^0e^{-i\omega_0 s}\to 0$, as $s\to \infty$,
and write the corresponding nonstationary solutions in the form
$$
u(y,t)=A(y,t)e^{-i\omega_0t}.
$$
We prove that $A(y,t)\to A_\infty(y)$
as $t\to\infty$, where $A_\infty(y)$ is a solution to the corresponding stationary
Helmholtz equation.

The key role in this
asymptotic analysis plays the Sommerfeld-Malujinetz type representation
for the diffracted wave
\begin{equation}\label{ud}
u_{d}(\rho,\theta,t):=\ds\frac{i}{4\Phi}\int_{\R}Z(\beta+i\theta)F(t-\rho\cosh\beta)d\beta,~~\theta\in \Theta:=
[\phi,2\pi]\setminus \{\theta_1,\theta_2\}
\end{equation}
in the case of locally summable incident wave.
We will use it for an analysis of
asymptotic behavior of the diffracted wave near the wave front
and for large times.
The representation  was obtained first in \cite{kmm}-\cite{eam}
for the Heaviside-type smooth incident wave (\ref{uin}) using the method of
complex characteristics \cite{3}-\cite{kmr}.
Here we extend this representation to locally summable incident waves.
The representation was used
in \cite {cmkz}, and \cite {eam} to find the convergence rate
to the limiting amplitude.

Let us comment on previous works.
The scattering by wedge of
incident wave (\ref{dia})
was considered for the first time by Sobolev \cite{Sob34} - \cite{Sob1}
in 1930', by Keller and Blank \cite {KB} in 1951,
by Kay  \cite {Ka} in 1953 and Rottbrand \cite{Ro,Rot} in 1998.

The problem obviously reduces to the
Heaviside incident wave $F(s)=h(s)$.
For this step function
Sobolev construct in  \cite{Sob34}
the solution in the form
\be\la{sf}
u(y,t)=g(\zeta(y,t)).
\ee
Here $\zeta(y,t)$ is an ``algebraic'' function defined by equation
\be\la{zet}
bt-m(\zeta)y_1-n(\zeta)y_2-\chi(\zeta)=0,
\ee
where $m(\zeta)$, $n(\zeta)$ and $\chi(\zeta)$ are suitable complex
analytic functions related by
\be\la{mn}
m^2(\zeta)+n^2(\zeta)=1.
\ee
Sobolev refers
the formula (\re{sf}) as the Sobolev-Smirnov representation
and relates it to a dilation
invariance of the wave equation.

The problem is solved explicitly
using conformal mappings
onto unit circle
and Schwarz's reflection principle:
antisymmetric reflections in the DD case,
and
symmetric reflections in the NN case.

The resulting formula coincides with our formula
(\re{svertka-i})
as we will
prove in this paper.

In the next paper \ci{Sob}
(mainly included in  \ci{Sob1})
Sobolev relates
this process of the reflections with the wave
propagation on the logarithmic Riemann surface which is
in spirit of the Sommerfeld ideas cited in \ci{Sob1}.

In these papers
Sobolev introduced the famous discontinuous
``weak solutions'' to the wave equations
which appear the cornerstone for the Theory
of Distributions developed later by L. Schwartz.
\medskip

Keller and Blank \cite{KB} also considered
diffraction of the Heaviside incident wave by a wedge
developing Busemann's ``Conical Flow Method''
which is in the same spirit as
the Sobolev approach:  the dilation
invariance of the wave equation
allows to reduce the problem to
the  Laplace equation on a circle with
piecewise constant boundary values. The obtained solution coincides with the Sobolev formula
(and with our solution) as we will prove elsewhere.
\medskip

Kay's  approach \cite {Ka} relies
on  separation of variables in the wave equation in
suitable variables.
Any
solution  of the wave equation is represented
in the form of series in the Whittaker functions \cite[p. 279]{WW}.
The author proves that the series
coincide with the Keller-Blank solution
in the case of  the Heaviside incident wave
(see p.  434 of \cite {Ka}).
\medskip

Rottbrand \cite{Ro,Rot}
considered the diffraction of the plane wave
 (\ref{dia}) with
$F(s)=\ds \int_0^s g(\tau)d\tau$
where $g\in L^1(\R)$,  $\supp g\subset(0,\infty)$.

The problem is reduced by a conformal map
to the  Rawlins's mixed problem
which is solved in \cite{Ro} using the Wiener-Hopf technique.

The solution for the incident wave with
$F(s)=s^{-1/2}_+$  has been
constructed by Borovikov \cite{bor} who  used this formula
to reproduce the solution of Sobolev.
\medskip

The formulas obtained in
\ci{Sob34}-\ci{Sob1} and \cite{KB}, \cite{Ka}, and \cite{Ro}
appear quite different.
It is instructive to note
that in all these works
the classes of solutions are not specified, and
the uniqueness of solutions is not analyzed.

In our paper we construct the  solution
in a suitable  space of distributions
for incident wave (\ref{dia}) with any tempered
distribution $F$ with the support in $\ov\R$.
Moreover, we
prove that the solution is unique in this class,
and is given by convolution (\re{svertka-i}).
Let us stress, that we deduce the existence and uniqueness
of solutions
from our previous results \ci{kmr,kmm}.

We obtain the  Sobolev formula
for the theta-function incident wave.
This justification of the diffraction formula
was one of our main motivation
in writing this paper.
The coincidence with the Keller-Blank formula requires
more calculations and will be published
elsewhere.

\medskip

Let us outline the plan of our paper. In Section 2 we reduce
the problem (\ref{sidlu}) by the Fourier-Laplace transform.
In Section 3 we obtain the convolution formula for the solution
to (\ref{sidlu}) . In Sections 4 and 5 we study asymptotics of
the solutions as $t\to\infty$. In Section 6 we check that our
general formula coincides with the Sobolev result for the Heaviside function.
In Appendix we calculate some Fourier transforms.


\section{Formulation of the scattering problem}
\setcounter{equation}{0}
The front of the incident wave $u_{\rm in}(y,t)$ at any moment of time $t\leq 0$
is the straight line $\big\{y:t-n_{0}\cdot y=0\big\}$ in $\R^2$.
For $n_{0}\cdot y>t$, we have $u_{\rm in}(y,t)=0$ by
(\ref{F}).  We impose the following conditions on the vector $n_{0}$.
First, we suppose that
 $\phi-\pi/2<\alpha<\pi/2$. Then
the front of  $u_{\rm in}(y,t)$
lies in $Q$ for  $t< 0$.

Second, we suppose that  the incident wave is reflected by both sides of the wedge.
This is equivalent to the condition $0<\alpha<\phi$.
These two conditions on the vector $n_0$ are expressed by the inequalities:
 \begin{equation}\label{ron}
 \max ( 0,\phi-\pi/2)
           <\alpha<\min( \pi/2 ,\phi).
 \end{equation}
The extension of our results to another angles $\phi$ and $\alpha$
does not pose any new difficulties. In particular, the formulas
(\ref{ud})-(\ref{hih}) remain valid for all angles $\Phi$ and $\alpha$
(see Figure 1).
\\\\\\\\\\\\\\\\\\\\\\
\setlength{\unitlength}{1mm}
 \begin{picture} (110,50)
 \put(100,35) {\vector (1,0){39}} \put(138,37){$y_1$}
  \put (55,35){\line(1,0){50}}

 \put
 (55,35){\vector(0,1){50}} \put(57,85){$y_2$}




\multiput(55,35)(5,4){12}{\line(5,4){4}}


\put(116,83){$\theta=\theta_1$}


 \qbezier(68,38.5)(68,40)(67,41)

 \qbezier(69,38.7)(69,40)(68,41.6)

 \qbezier(67,41)(67.3,42.5)(66,44)

 \qbezier(68, 41.6)(68.71,42.5)(66.9,44.8)


\multiput(55,35)(5,-1.25){13}{\line(4,-1){4}}
\put(120,17){$\theta=\theta_2$}

 \qbezier(65,35)(65.6,34)(64.7, 32.9)


  \put(55,35){\line(2,1){65}}


  \put (105,45){W}


  \put (20,55){Q}

  \put(4,47){\line(1,-4){10}}

  \put(-3,45){\line(1,-4){10}}

  \put(11,49){\line(1,-4){10}}

  \put(19,51){\line(1,-4){10}}

  \put(26,53){\line(1,-4){10}}

  \put(34,55){\line(1,-4){10}}

  \put(42,57){\line(1,-4){10}}

  \put(55,31){0}

  \put (3,22){\line(4,1){60}}


  \put(55,35) {\vector (4,1){25}}

 \qbezier(65,35)(65.6,36)(64.7,37.3)


 \put(67,35.5){$\alpha$}

 \qbezier(95,35)(98,39)(96,44) \qbezier(96,44)(95,46)(94,48)
 \qbezier(94,48)(95,46)(94,48) \qbezier(94,48)(93,49)(92,50)
 \qbezier(92,50)(91,51)(90,52)
 \qbezier(90,52)(89.5,52.4)(89.5,52.4)
  \put(81,41){$\vec{n_0}$}

 \put (98,47){$\phi$}

 \put (120,31){$Q_1$}


 \put (115,70){$Q_2$}



  \qbezier(3,22)(10,40)(20,20)
 \qbezier(20,20)(25,10)(30,26)
 \qbezier(30,26)(35,40)(43,35)
 \qbezier(43,35)(50,30)(55,35)


 \end{picture}

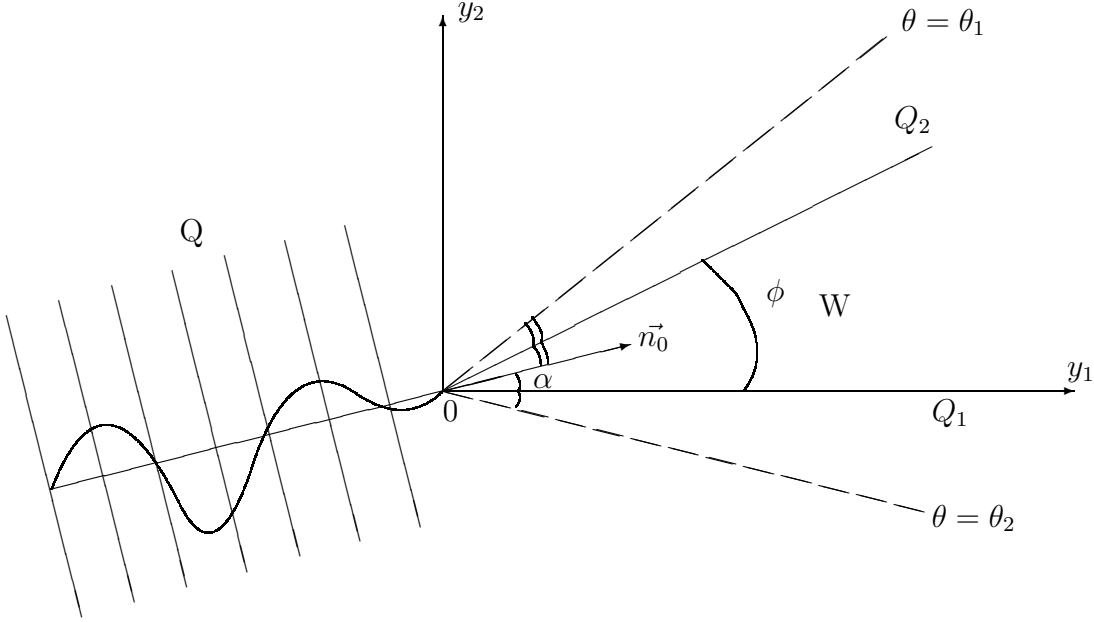
 \centerline{Figure 1. The incident plane wave}
\medskip
\medskip
Let us denote by $u(y,t)$ a solution of problem (\ref{sidlu}) and by
\begin{equation}\label{ssi}
u_{s}(y,t):=u(y,t)-u_{\rm in}(y,t)
\end{equation}
the scattered wave. Then $u_s$
is a solution to the following mixed problem:
\begin{equation}\label{sidlus}
\left\{
\ba{l}
\Box u_s(y,t)=0,~y\in Q,~~B u_s(y,t)\vert_{Q_1\cup Q_2}=-
B u_{\rm in}(y,t)\vert_{Q_1\cup Q_2},~t\in \R,\\
~~u_s(y,t)=0,~y\in Q,~t<0.
\end{array}
\right.
\end{equation}
Let us define  the meaning of this mixed problem. For a function  $u(t)\in
S(\R)$ we denote its Fourier transform
 \begin{equation}\label{1.80}
 \hat u(\omega):=F_{t\to \omega} u(\omega):=\int_{\R} e^{i\omega t}u(t)dt,~~\omega\in \R.
\end{equation}
This transform is extended
by the continuity to  tempered distributions $u\in
S'(\R)$.
For the case  $\supp
 u\subset \overline{\R^+}$ the distribution $\hat u(\omega)$ admits  an analytic
extension to the upper half plane  $\C^+:=\{z\in \C:~ \Im z>0\}$ and
\be\la{pw}
|\hat f(\omega)|\leq C(1+|\omega|)^m|\Im \omega|^{-N},~~\omega\in \C^+
\ee
for  some $m,N\geq 0$ by the Paley-Wiener Theorem.
We will call this analytic continuation as the Fourier-Laplace transform of $f$.
Conversely, if an analytic function $G(\omega)$
in $\C^+$ satisfies
(\ref{pw}) then there exists its boundary value
as $\Im \omega\to 0+$ in the sense of $S'(\R)$, see \ci[Thm I.5.2]{41}.

\no Let us  introduce functional spaces of solutions. First we define spaces of test functions.
For $\varphi(y,t)\in C^{\infty}(\ov Q\times \R)$ let us denote
\be\la{mN}
\|\varphi\|_{m,N}=\sup
\limits_{(y,t)\in \ov Q\times \R,~|\alpha|\leq m}
(1+|y|+|t|)^N|\partial ^\alpha_{y,t}\varphi(y,t)|
\ee
Similarly, for $\varphi(y)\in C^{\infty}(\ov Q)$ let us denote
\be\la{mN2}
\|\varphi\|_{m,N}=\sup\limits_{y\in \ov Q,~~|\alpha|\leq m}(1+|y|)^N|\partial ^\alpha_{y}\varphi(y)|
\ee
\begin{definition} We denote the countably-normed spaces:
\\
i) $S(\ov Q\times \R):=\{
\varphi(y,t)\in C^{\infty}(\ov Q\times \R):~~
\|\varphi\|_{m,N}<\infty,~~~m,N>0
\}.
$
\\
ii) $
S(\ov Q):=\{
\varphi(y)\in C^{\infty}(\ov Q)\}:~~\|\varphi\|_{m,N}<\infty,~~~m,N>0\}.
$
\end{definition}
Now we define the space $S'(\ov Q \times {\ov \R^+})$
of tempered distributions with supports in ${\ov Q\times \ov {\R^+}}$:
\begin{definition}\label{dist}
$S'(\ov Q\times {\ov \R^+})$ is
the space of  linear continuous functionals on $S(\ov Q\times \R)$ with supports
in $\ov Q\times \ov {\R^+}$.
\end{definition}
For each $u\in S'(\ov Q\times \ov{\R^+})$ there exist $m, N \geq 0$ such that
\be\la{ocpq}
|\langle u(y,t),\varphi(y,t)\rangle|\leq C\|\varphi\|_{m,N},
~~\varphi\in S(\ov Q\times \R)
\ee
This follows  from definition of the topology in
countable-normed spaces as noted in
\ci[Ch. I, \S4]{GS}.
We will use the following  Paley-Wiener Theorem for distributions
which is a straightforward generalization of \ci[Thm I.5.2]{41}.
\begin{lemma}\la{lpewidistr}
(i) Let $u\in S'(\ov Q\times\ov{\R^+})$.
Then its Fourier transform
$\hat u(y,\omega)$ extends to an  analytic  function on $\C^+$ with values in $S'(\ov Q)$, and
there exist $m,N\geq 0$ s.t.
\be\la{hp}
\langle \hat u(y,\omega),\varphi(y)\rangle|\leq C\|\varphi\|_{m,N}(1+|\omega|)^m |\Im\omega|^{-N},~~ \varphi\in S(\ov Q)
\ee
(ii) Conversely, let $\hat u(y,\omega)$  be an analytic function of $\omega\in \C^+$
with values in $S'(\ov Q)$ and the bound
(\ref{hp}) holds for some  $m,N\geq 0$.
Then $\hat u(y,\omega)$ is the Fourier-Laplace  transform of
a distribution $u\in S'(\ov Q\times \ov{ \R^+})$.
\end{lemma}
\begin{definition}\la{dhp}
We denote by $HP(\C^+,S'(\ov Q))$
the space of holomorphic functions in
$\C^+$ with values in $S'(\ov Q)$
satisfying the bound (\ref{hp}) for some $m,N$.

\end{definition}
Let us introduce the space of solutions  to (\ref{sidlus}).
\begin{definition}
\label{dmeps}(see Def 2.1 \cite {kmm})
(i) $E_\ve$ is the Banach
space of functions $u(y)\in C(\ov Q)\cap C^1(\dot {\ov Q})$ with finite norm
\begin{equation}\label{eeps}
\|u\|_{\ve}= \sup \limits_{y\in \ov Q}|u(y)|+\sup \limits_{y\in \dot {\ov Q}}\{y\}^{\ve}|\nabla u(y)|<\infty
\end{equation}
where $\{y\}:=\frac{|y|}{1+|y|}$ and $\dot {\ov{Q}}:=\ov {Q}\setminus 0.$

\no (ii) ${\cal M}_\ve$ is the space of
tempered distributions $u(y,t)\in S'(\ov Q\times \ov{\R^+})$, such that its
Fourier-Laplace  transform ${\hat u}(y,\omega)$ is a holomorphic function of $\omega \in \C^+$ with the values in $E_\ve$.
\end{definition}
For $u\in {\cal M}_\ve$
the Fourier transform in the system (\ref{sidlus})
gives
\begin{eqnarray}\label{stdd}
\left\{
\begin{array}{l}
 (-\Delta-\omega^2) \hat u_s(y,\omega)=0~~y\in Q\\
 \displaystyle \hat u_s(y,\omega)
 =-\hat F(\omega)e^{i\omega y_1 \cos\alpha },~~y\in Q_1\\
 \displaystyle \hat u_s(y,\omega)=- \hat F(\omega)\displaystyle
 e^{-i\omega y_2 [\cos(\alpha+\Phi)/\sin\Phi]},
 ~~y\in Q_2\\
 \end{array}
 \right |
 \omega\in \C^+
 \end{eqnarray}
in the case of DD-problem, and similar equations hold for NN and DN -problems (see Appendix A1).

Let us note that the boundary conditions in (\ref{stdd})  are well defined for $\hat u_s(y,\omega)\in E_{\ve}$ in contrast to
the boundary conditions in (\ref{sidlus}) which are not well defined for tempered distributions
$u_s(y,t)$. This suggests the following definition.
\begin{definition}\label{kakponsi}
We call an $u_s(y,t)\in \cal M_\ve$ a solution to (\ref{sidlus}) if
$\hat u_s(y,\omega)$ is the solution to (\ref{stdd}).
\end{definition}


\section{Existence and uniqueness}
\setcounter{equation}{0}

In this section we prove the uniqueness and existence of solution to the scattering problem
(\ref{sidlu}) in class $\cal M_\ve$, using methods and results of \cite {kmm}-\cite {eam}.
We will assume  that
\be\la{fd}
F\in S'(\R),~~{\rm supp}~ F\subset \ov{ \R^+}.
\ee
We will prove the existence and uniqueness of a solution to  the problem (\ref{sidlus})
with any fixed boundary operators  $B _1$ and  $B_2$.

\subsection{Uniqueness}

\begin{theorem}
A solution to problem (\ref{sidlus})
is unique in the class $\cal M_{\ve}$ for any $\ve\in (0,1)$.
\end{theorem}
{\bf Proof.} Let $u_s(y,t)\in {\cal M}_\ve$ satisfy $(\ref{sidlus})$.
By definition (\ref{kakponsi}) it suffices to prove the
uniqueness of solution $\hat u(y,\omega)$ to problems (\ref{stdd}) for any $\omega \in \C^+$ in the space
$E_{\ve}$. This uniqueness is proved in Sections 7 and 8 of \cite {kmm}
for DD-problem, in \cite{MM1}, \cite{mm} for DN-problem and in \cite{eam} for NN-problem. \bo
\medskip

\subsection{Existence}
Let us recall the functions ${\cal S}_s(y,\omega)$,
${\cal S}_d(y,\omega)$, ${\cal S}_r(y,\omega)$ introduced in \cite {km}, \cite{mm}
and in \cite{eam} for DD, DN and NN-problems respectively
which are the densities of scattered, diffracted and reflected waves respectively and
\begin{eqnarray}\label{hatjdjr}
\left.
\ba{rcl}
{\cal S}_r(\rho,\theta,\omega)&:=&
\left\{
\begin{array}{ll}
-e^{i\omega\rho\cos(\theta-\theta_1)},&\phi< \theta< \theta_1\\
0,&\theta_1 <\theta<\theta_2\\
- e^{i\omega\rho\cos(\theta-\theta_2)},&\theta_2<\theta<
2\pi
\end{array}
\right.
\\
\\
{\cal S}_d(\rho,\theta,\omega)&:=&\displaystyle\frac{i}{4\Phi}\int_{\R}
e^{i\omega\rho\cosh\beta}Z(\beta+i\theta)d\beta,~~\theta\not=\theta_{1,2}\\
\\
{\cal S}_s(\rho,\theta,\omega)&:=&{\cal S}_r(\rho,\theta,\omega)+{\cal S}_d(\rho,\theta,\omega),~~
\theta\not=\theta_{1,2}
\ea
\right| \rho>0,~~~\omega\in\overline{\C^+}.
\end{eqnarray}
Here
\be\la{teta12}
\theta_1:=2\phi-\alpha,~~~ \theta_2:=2\pi-\alpha,
\ee
are the ``critical" directions,
and
\begin{equation}\label{z}
Z(\beta)=-H(-\frac{i\pi}{2}+\beta)+H(-\frac{5i\pi}{2}+\beta),~~\beta\in \C,
\end{equation}
where $H$ is  the Malyuzhinets
type kernels $H$ for DD, NN and DN-problems (see Appendix A2).
The formulas for ${\cal S}_r$ for the DN and NN-problems are given in the Appendix A3.

By \cite [Thm 8.1] {km}, the function
${\cal S}_s(y,\omega)\in C_b(\ov Q\times \C^+)$,
and it is analytic in $\omega\in \C^+$. This implies that
\be\la{ss}
{\cal S}_s\in HP(\C^+,S'(\ov Q)).
\ee
\begin{equation}\label{i98}
{\cal Z}(\beta):=Z(\beta)+Z(-\beta),~~
l(\lambda)=\left\{
\begin{array}{ll}
\ln(\lambda+\sqrt{\lambda^2-1}),~~~\lambda\geq1\\\\
 0~~~~~~~~~~~~~~~~~~~~,~~~\lambda\in(0,1);
\end{array}
 \right.
 \end{equation}

In  Appendix A4 we calculate the inverse Fourier transforms
$F_{\om\to t}$
of ${\cal S}_r$ and ${\cal S}_d$ which we denote by
${\cal J}_r(\rho,\theta,t)$, and $ {\cal J}_d(\rho,\theta,t)$ respectively:
\beqn\la{J}
\left.
\ba{rcl}
{\cal J}_r(\rho,\theta,t)&=&
\left\{
\ba{ll}
\displaystyle
\delta(t-\rho\cos(\theta-\theta_1)),&\phi<\theta<\theta_1\\
0, & \theta_1<\theta<\theta_2\\
\delta(t-\rho\cos(\theta-\theta_2)),&\theta_2<\theta<2\pi\\
\ea
\right.,
\\
\\
{\cal J}_d(\rho,\theta,t)&=&
\ds
\ba{ll}
\displaystyle
\frac{i}{4\Phi}\frac{{\cal Z}(l(t/\rho)+i\theta)}{\sqrt{t^2-\rho^2}}h(t-\rho)&
\ea
\\
\\
{\cal J}_s(\rho,\theta,t)&=&{\cal J}_r(\rho,\theta,t)+{\cal J}_d(\rho,\theta,t),~~
\theta\not=\theta_{1,2}
\ea
\right|~~~\rho>0, ~~~t\in\R,
\eeqn
where $h(\cdot)$ denotes the Heaviside function.
Let us note that ${\cal J}_r(\rho,\theta,t)={\cal J}_d(\rho,\theta,t)=0$
for $t<0$.
For the NN and DN-problems the functions ${\cal J}_r$ and ${\cal J}_d$,
are calculated in Appendix A5.

For our application it is crucially important that
\be\la{jsp}
{\cal J}_d,~{\cal J}_r~{\cal J}_s\in S'({\ov Q}\times\ov {\R^+}).
\ee
This follows immediately from Lemma \ref{lpewidistr}, (\ref{ss})
and the fact that ${\cal J}_s$ is the inverse Fourier transform of ${\cal S}_s$.

Now we prove  main  theorem on existence of solution to (\ref{sidlus}) and its convolution representation.

By Definition \ref{kakponsi} the
problem (\ref{sidlus}) is equivalent
to (\ref{stdd}).
Let us recall basic results of
\cite{kmm}--\cite{eam}.
\begin{lemma}\la{lsmooth}
Let
 the incident wave (\ref{uin}) corresponds to
the smooth profile function (\re{f}).
Then
\medskip\\
i)
The unique solution  $\hat u_s(y,\om)\in
E_\ve$
to (\ref{stdd})
is given by
\be\label{clresh}
\hat u_s(\rho,\theta,\omega)=\hat F(\omega){\cal S}_s(\rho,\theta,\omega),~~~\omega\in \C^+.
\ee
where $\hat F(\om)=\hat f(\omega-\omega_0)$.
\medskip\\
ii) The parameter $\ve$ is given by
\be\label{ve}
\ve=
\left\{
\ba{ll}
1-\ds\frac\pi{2\Phi}&\mbox{\rm for DD and NN cases,}\\
\\
1-\ds\frac\pi{\Phi}&\mbox{\rm for DN case.}\\
\ea\right.
\ee
\end{lemma}

The formula (\re{clresh}) is proved in \ci[(3.15)]{km}, while
(\re{ve}) are found
 in Section 10 of \cite{km} for the DD-problem,
 in Section 6  of \cite{eam} for the NN-problem and
in  Section 16 of \cite{mm} for the DN-problem.
This lemma implies, in particular, that
\be\la{usE}
{\cal S}_s(\cdot,\omega) \in E_\ve,~ \omega\in \C^+,
\ee
since for any $\omega\in \C^+$
we can choose a smooth profile function (\re{f}) such that
$ \hat f(\omega-\omega_0)\not = 0$.
\begin{corollary}\label{cr}
The function ${\cal S}_s(y,\omega)$ is a solution to problem (\ref{stdd}) with $\hat F\equiv 1$.
\end{corollary}

Our main result is the following theorem.

\begin{theorem}\label{sush}
Let $F$ satisfy (\ref{fd}). Then
\medskip\\\
i)
There exists a generalized solution
$u_s(y,t)\in
{\cal M}_\ve$ to problem (\ref{sidlus})
with $\ve$ given by (\re{ve}).
\medskip\\\
ii)
The solution is given by the convolution
\be\label{svertka}
u_s=F_\de*{\cal J}_s,~~
(y,t)\in \overline Q\times  \R,
\ee
where $F_\de(y,t):=F(t)\de (y)$, and the convolution is well defined
in the sense of distributions.

\end{theorem}
{\bf Proof.} i)
For any distribution (\ref{fd})
it is natural to  define the solution
to  (\ref{stdd})
again by (\ref{clresh}).
Indeed, $\hat u_s(\cdot,\omega)\in E_\ve$ for $\omega\in \C^+$ since
${\cal S}_s(\cdot,\omega)\in E_\ve$
by (\ref{usE}).
Moreover, $\hat u_s$ is a solution to (\ref{stdd})
by Corollary \ref{cr}.
It remains to prove that
\be\la{us2}
u_s(y,t):=F^{-1}_{\om\to t}
\hat  u_s(y,\omega)
\in {\cal M}_\ve.
\ee
It suffices to check that $u_s\in S'(\ov Q)\times \ov {\R^+}$,
or equivalently,
$\hat u_s\in HP(\C^+,S'(\ov Q))$.
First, we note that $\hat F\in HP(\C^+,S'(\ov Q))$ by Paley-Wiener Theorem
\cite [Thm I.5.2] {41}. Second, ${\cal S}_s\in HP( {\C^+},S'(\ov Q))$
by (\ref{ss}). Hence, the product (\ref{clresh})
also belongs to $HP(\C^+,S'(\ov Q))$ since $HP$ is algebra.

ii) The convolution representation (\ref{svertka}) follows from
 (\ref{clresh}).
The convolution is well defined since
the intersection of the  supports of $F_{\delta} (y',t')$ and
${\cal J}_s (y-y',t-t')$ is a bounded set
for any fixed $y\in \ov Q$ and $t\in \R$. \bo
\medskip

\subsection{Sommerfeld type representation of diffracted wave}

Let us substitute  the splitting from the
last line of  (\ref{J}) into
(\ref{svertka}). Then we obtain the corresponding splitting
\be\la{us}
u_s=u_r+u_d.
\ee
By (\ref{clresh}), we obtain for the DD-case
\be\la{udur}
u_r=F^{-1}_{\omega\to t} [\hat F {\cal S}_r]=F_{\delta}*{\cal J}_r,~~
u_d=F^{-1}_{\omega\to t} [\hat F  {\cal S}_d]
=F_{\delta}*{\cal J}_d.
\ee
Similar formulas hold for
NN and DN-cases. The explicit expressions of $u_r$ for all types of boundary conditions are given in Appendix A6.
\begin{lemma}\label{lpr/ud}
Suppose that
\be\la{fs}
F\in L^1_{loc}(\R)
\ee
and (\ref{F}) holds.
Then the diffracted wave
 $u_d$ admits the representation (\ref{ud})
with a suitable kernel $Z$
for any boundary conditions of the DD, NN or DN-types.
\end{lemma}
{\bf Proof.} It suffice to prove that
\be\la{FS}
\hat u_d
(\rho,\theta,\om)
=
{\hat F}(\om)  {\cal S}_d(\rho,\theta,\om)
=\frac{i}{4\Phi}\int_{\R} e^{i\omega t}
\left (\int_{\R} Z(\beta+i\theta)F(t-\rho\cosh \beta)d\beta\right )dt,~~\omega\in \C^+.
\ee


Let us denote
\be\la{q}
q=\fr \pi{2\Phi}.
\ee
>From (\ref{z}) and (\ref{zzic}), (\ref{hih}) we get the decay
\begin{equation}\label{ocz}
|Z(\beta+i\theta)|\leq C(\theta)e^{-2q|\beta|}~~,~~~~~
\theta\in\Theta
\end{equation}
for the $DD$- and  $NN$ problems, and
\begin{equation}\label{ocz1}
|Z(\beta+i\theta)|\leq C(\theta)e^{-q|\beta|}~~,~~~~~\theta\in\Theta
\end{equation}
for the $DN$-problem. Hence (\ref{F}) implies by the Fubini Theorem,
\be\la{Fub}
\ba{l}\ds
\frac{i}{4\Phi}\int_{\R} e^{i\omega t}\left (\int_{\R} Z(\beta+i\theta)F(t-\rho\cosh \beta)d\beta\right )dt\\\ds =
\frac{i}{4\Phi}\int_{\R}  Z(\beta+i\theta)\left (
\int_{\R} e^{i\omega t} F(t-\rho\cosh \beta)dt\right )d\beta\ds=
{\hat F}(\om)  {\cal S}_d(\rho,\theta,\om)
,~~\omega\in \C^+,
\ea
\ee
by formula (\ref{hatjdjr}) for ${\cal S}_d$.
\bo
\section{Stabilization of the diffracted wave}
\setcounter{equation}{0}

Let $l_k$ be the critical rays $l_k:=\{(\rho,\theta_k): ~~\rho>0\}$, $k=1,2$.
\begin{lemma}\label{lskachok}
Let (\ref{ron}), (\ref{fs}) and (\ref{F}) hold.  Then for any type of  boundary
conditions
(DD, NN and DN) and $t\in\R$
there exist the limits
$$
u_d(\rho,\theta_k\pm 0,t):=
\lim\limits_{\ve\to 0+}u_d(\rho,\theta_k\pm \ve,t),~~
\rho>0, ~~k=1,2
$$
in the sense of distribution, and the jumps
of $u_d$ on the critical rays
for the DD-problem
are given by
\be\label{jk}
[u_d]_k(\rho,t):= u(\rho,\theta_k+0,t)- u(\rho,\theta_k-0,t)=(-1)^{k+1}F(t-\rho),
~~~\rho>0,~k=1,2.
\ee

\end{lemma}
\no {\bf Proof.} We will use the representation (\ref{ud}) and we
will consider DD case for the concreteness. The cases of NN and DN-problems are analyzed similarly
(see Appendix A6). Formulas (\ref{z}) and (\ref{zzic}) imply the following representation:
\begin{equation}\label{n2}
\begin{array}{ll}
Z(\beta+i\theta)=-\coth\Big(q\beta+ic_0\Big)\mp\coth\Big(q\beta+ic_1\Big)
\pm\coth\Big(q\beta+ic_2\Big)+\coth\Big(q\beta+ic_3\Big)
\end{array}
\end{equation}
for DD and NN-cases respectively,
where
\begin{equation}\label{9ww}
c_k:=q(\theta-p_k);~~~p_0= \alpha, p_1=\theta_1, p_2=\theta_2, p_3=2\pi+\alpha.
%
\end{equation}
First let us consider the case when
$F$ is a smooth  function
satisfying (\ref{F}).
Then the  Sokhotski-Plemelj formulas
imply
$$
\ba{l}
[u_d]_k(\rho,t)=
\ds\int_{-1}^{1}
F(t-\rho\cosh\beta)\Big[\coth(q\beta+i0)-
\coth(q\beta-i0)\Big]d\beta=(-1)^{k+1}F(t-\rho)
\ea
$$
since $\coth(q\beta+ic_k)$
with $k=0$ and $k=3$
are continuous
on the critical rays
for  $\alpha$ satisfying (\ref{ron}).
For $F$ satisfying (\ref{F}),
(\ref{jk}) holds in the sense of distributions. \bo
\medskip

\begin{theorem}\label{tstud}
Let the incident wave profile $F$ satisfy (\ref{F}), and
\be\la{C}
F(s)\to C,~~~~~~~s\to\infty.
\ee
Then
\medskip\\
i) The diffracted wave converges in the long time limit:
\begin{equation}\label{vtr6}
\begin{array}{c}
u_{d}(\rho,\theta,t)~_{\overrightarrow{t\to\infty}}~
u_d(\theta,\infty):=\ds\frac{iC}{4\Phi}\int_{\R}Z(\beta+i\theta)d\beta,
~~\rho>0,~~\theta\in \Theta,
\\
\end{array}
\end{equation}
and in particular,
\be\la{udk}
[u_d]_k(\rho,t)~_{\overrightarrow{t\to\infty}}~C,
~~~~
\mbox{\rm for}~~~\rho>0,~~ k=1,2.
\ee
ii) Conversely, (\re{udk}) implies (\re{C}).

\end{theorem}
\no {\bf Proof.} We can use (\ref{ud}) by Lemma \ref{lpr/ud}.
\medskip\\
i)
Conditions (\ref{F}) and (\ref{ud}) imply that
\begin{equation}\label{ud/kon}
u_{d}(\rho,\theta,t)=\ds\frac{i}{4\Phi}\int_{-l(t/\rho)}^{l(t/\rho)}Z(\beta+i\theta)F(t-\rho\cosh\beta)d\beta,~~\theta\in \Theta
\end{equation}
where $l(\cdot)$ is defined by (\ref{i98}).  Then (\ref{vtr6}) follows from (\ref{ocz} by the Lebesgue Dominate Convergence Theorem.
The convergence (\re{udk}) follows from (\ref{jk}) and (\re{C}).
\medskip\\
ii) (\re{C}) follows from (\ref{jk}) and (\re{udk}).   \bo

\begin{corollary}
For any type of bpundary conditions (DD, NN or DN)
the function $u_d(\theta,\infty)$ is the piecewise constant function
of $\theta\in \Theta$ with the jumps at
$\theta=\theta_1$ and $\theta=\theta_2$.
\end{corollary}
{\bf Proof.} For the DD and NN cases formula (\ref{n2}) implies that
$Z(\beta)$ is holomorphic on $\C\setminus P$ where
$P=\cup_{l=0,1,2,3} \{ip_l+2ik\Phi  :~~k\in \Z\}$.
Moreover, $p_0,p_3\not\in \Theta$  by (\ref{ron}).
Hence,
$Z(\beta+i\theta)$ may have a pole $\beta\in\R$ only at
 $\beta=0$, and it holds only for $\theta=\theta_1$ or $\theta=\theta_2$.
Therefore, the corollary follows from the decay (\ref{ocz}) and the Cauchy Theorem.

For the DN-case the proof is similar relying on the decay (\ref{ocz1}). \bo
\medskip

\section{Limiting Amplitude Principle}
\setcounter{equation}{0}
Consider the incident wave
$$
F^0(t):=a^0 e^{-i\omega_0 t}h(t),~~t\in \R
$$
where $\omega_0\not = 0$ (the case $\omega_0=0$
is covered by Theorem (\ref{tstud})). By (\ref{ud}) the corresponding
diffracted wave
is given by
$$
u_d^0(\rho,\theta,t)=
i\frac{e^{-i\omega_0t}}{4\Phi}\int_{-l(t/\rho)}^{l(t/\rho)}e^{i\omega\rho\cosh \beta}Z(\beta+i\theta)d\beta
$$
where $Z$ is given by
(\ref{n2})
for the DD- and NN
 problems (and
by (\ref{hih}) for the DN-problem),
and $l(\cdot)$ is defined by (\ref{i98}).
The limiting amplitude of this wave is
\be\la{a0}
A^0(\rho,\theta)=\frac{i}{4\Phi}\int_{\R} a^0 e^{i\omega_0\rho\cosh \beta}Z(\beta+i\theta)d\beta.~~\forall \theta\in \Theta
\ee
since $l(t/\rho)\to \infty$, as $t\to \infty$ while $Z$ satisfies (\ref{ocz}) for DD and NN-problems and (\ref{ocz1}) for DN-problem.

Let us deine the amplitude $A_d(\rho,\theta, t)$ of the diffracted wave (\ref{ud}) by
\be\la{ad}
A_d(\rho,\theta, t):=\ds e^{i\omega_0t}\frac{i}{4\Phi}\int_{-l(t/\rho)}^{l(t/\rho)}Z(\beta+i\theta)F(t-\rho\cosh\beta)d\beta,
~~\rho>0,
~~\theta\in \Theta,~~t>0.
\ee
In the following theorem we prove that  the  amplitude is  asymptotically close to the amplitude
(\ref{a0}) if $F$ is asymptotically close to $F^0$.

\begin{theorem}\la{lap} {\rm (Limiting Amplitude Principle)}
Suppose that
\be\la{R}
R(t):=F(t)-F^0(t)\to 0,~~ t\to\infty.
\ee
Then for any $\de>0$
$$
A_d(\rho,\theta,t)-A
_0(\rho,\theta)\to 0,~~t\to\infty,
$$
uniformly in bounded $\rho>0$ and $\theta-\theta_k\ge \de$.

\end{theorem}
{\bf Proof.} By definitions (\ref{a0}) and  (\ref{ad}),
$$
\ba{l}
\ds A_d(\rho,\theta,t)-A^0(\rho,\theta)
\\
=\ds
-\frac{i}{4\Phi}\int\limits_{|\beta|\geq l(t/\rho)}
F^0(-\rho\cosh \beta)
Z(\beta+i\theta)
d\beta
+\frac{i}{4\Phi}\int\limits_{-l(t/\rho)}^{l(t/\rho)} e^{i\omega_0t}
Z(\beta+i\theta)R(t-\rho\cosh\beta)d\beta.
\ea
$$
Estimates (\ref{ocz}), (\ref{ocz1}) imply that
$$
\frac{i}{4\Phi}\int\limits_{|\beta|\geq l(t/\rho)}
F^0(-\rho\cosh \beta)
Z(\beta+i\theta)
d\beta=\frac{i}{4\Phi}\int\limits_{|\beta|\geq l(t/\rho)}e^{i\omega_0\rho\cosh \beta}
Z(\beta+i\theta)
d\beta\to 0,~~t\to\infty
$$
uniformly in $\rho>0$ and $\theta\in\Theta$.
It remains to prove that
$$
R_1(\rho,\theta,t):=\int\limits_{-l(t/\rho)}^{l(t/\rho)}
e^{i\omega_0t}Z(\beta+i\theta)R(t-\rho\cosh\beta)d\beta\to 0,~~t\to\infty
$$
uniformly in bounded $\rho>0$ and $\theta-\theta_k\ge \de>0$.
First, (\ref{ocz}), (\ref{ocz1}) and (\ref{R}) imply that
for any $\ve>0$ there exists $\beta(\ve)$ s.t.
\be\la{d2}
\int_{|\beta|\geq \beta(\ve)}|Z(\beta+i\theta)R(t-\rho\cosh\beta)|d\beta<\ve/2
\ee
 uniformly in $\rho>0$ and $\theta\in\Theta$.
Second, (\ref{R}) implies that for
$0<\rho\le b<\infty$ and $\theta-\theta_k\ge \de>0$
there exists $t(\ve,\de,b)$ such that
\be\la{d1}
2 \beta(\ve)|Z(\beta+i\theta) R(t-\rho\cosh\beta)|<\ve/2,
~~|\beta|<\beta(\ve),~~\theta\in \Theta,
~~~t>t(\ve,\de,b).
\ee
Then for $0<\rho\le b<\infty$ and $t>t(\ve,\de,b)$ we have
$$
|R_1(\rho,\theta,t)|\leq \int_{|\beta|\leq \beta(\ve)}
|Z(\beta+i\theta)R(t-\rho\cosh\beta)|d\beta
+
\int_{|\beta|\geq \beta(\ve)}|Z(\beta+i\theta)R(t-\rho\cosh\beta)|d\beta\ds<\ve
$$
by (\ref{d1}) and (\ref{d2}). \bo


\section{Application to the Sobolev problem}
\setcounter{equation}{0}


In this section we reproduce the Sobolev
formula for dispersion of the $\theta$-function.
First, we calculate the diffracted wave for $F(s)=h(s)$.
Let us choose below the branch of $\ln$ with
\be\la{b}
\Im \ln z\in (-\pi,\pi)~,~~~~~~z\in\C\setminus (-\infty,0].
\ee

\begin{proposition}\label{mov}
Let $F(s)=\theta (s)$ and $t>\rho>0$. Then
\medskip\\
i)
for the $DD$ and $NN$-problems respectively
\begin{equation}\label{sd1}
\begin{array}{ll}
u_d(\rho,\theta,t)=\ds\frac{i}{2\pi}\Big[-\ln U_0\mp\ln U_1\pm\ln U_2+\ln U_3\Big]~,
~~~~\theta\in\Theta
\end{array}
\end{equation}
where
\begin{equation}\label{yx8}
U_{k}=\ds\frac{b^{q}e^{ic_{k}}-b^{-q}e^{-ic_{k}}}{-(b^{q}e^{-ic_{k}}-b^{-q}e^{ic_{k}})},~~~k=0,1,2,3;~~
b=\frac{t}{\rho}+\sqrt{\bigg(\frac{t}{\rho}\bigg)^2-1},
\end{equation}
and $c_k$ are given by (\re{9ww})
\medskip\\
ii)
 For the $DN$-problem
\begin{equation}\label{u77}
u(\rho,\theta,t)=\ds\frac{i}{2\pi}\Big[\ln V_0-\ln V_1+\ln V_2-\ln V_3\Big]
\end{equation}
where
\be\la{vk}
V_{k}=\ds\frac{(b^{q}e^{ic_{k}}+1)(b^{q}e^{-ic_{k}}-1)}{(b^{q}e^{ic_{k}}-1)(b^{q}e^{-ic_{k}}+1)},~~k=0,1,2,3.
\ee

\end{proposition}
\no {\bf Proof.} i)~
Formula (\re{ud}) with  $F(s)=\theta (s)$ gives
\beqn\label{udg}
u_{d}(\rho,\theta,t)&=&\ds\frac{i}{4\Phi}\int_{-l(t/\rho)}^{l(t/\rho)}
Z(\beta+i\theta)d\beta
=
\frac{i}{4\Phi}\sum_{k=0}^3 s_k\int_{-l(t/\rho)}^{l(t/\rho)}
\coth(q\beta+ic_k)
d\beta
\nonumber\\
\nonumber\\
&=&\frac{i}{2\Phi}\sum_{k=0}^3
s_k
\ln\Bigg[\frac{b^{q}e^{ic_k}-b^{-q}e^{-ic_k}}{-(b^{q}e^{-ic_k}-b^{-q}e^{ic_k})}\Bigg],~~~~s_k=\pm 1,
\eeqn
since $Z$ and $l(t/\rho)$ are defined by (\ref{n2}) and
(\ref{i98}) respectively,
$c_k\in (-\pi,0)\cup(0,\pi)$, and $b>1$.
Now  (\ref{sd1}) is proved.
\medskip\\
ii)
Similarly to (\re{udg}),
\beqn\label{uddn}
u_{d}(\rho,\theta,t)&=&\ds\frac{i}{4\Phi}\int_{-l(t/\rho)}^{l(t/\rho)}
Z(\beta+i\theta)d\beta
=
\frac{i}{4\Phi}\sum_{k=0}^3 s_k\int_{-l(t/\rho)}^{l(t/\rho)}
\frac{d\beta}{\sinh (q\beta+ic_k)}
\nonumber\\
\nonumber\\
&=&\frac{i}{4\Phi}\left (-\frac{1}{q}\right )\sum_{k=0}^3
s_k
\ln\Bigg[\frac{(b^{q}e^{ic_k}+1)(b^{q}e^{-ic_k}-1)}
{(b^{q}e^{ic_k}-1)(b^{q}e^{-ic_k}+1)}\Bigg],~~~~s_k=\pm 1,
\eeqn
since $Z$ is defined now by (\ref{n2})  and
$$
{\rm arg}~\ds\frac{b^{q}e^{ic}+1}{b^{q}e^{ic}-1}-{\rm arg}~\ds\frac{b^{q}e^{-ic}+1}{b^{q}e^{-ic}-1}\in(-\pi,\pi)
$$
where ${\rm arg}~(\cdot)\in(-\pi,\pi)$.
Now (\ref{u77}) is proved. \bo
\begin{corollary}
Let the incident wave $F$ be the Heaviside function.
Then the diffracted wave $u_d$ admits the following limits  as $t\to\infty$:
\medskip\\
i) for the $DD$-case,
\begin{equation}\label{kjl77}
u_d(\rho,\theta,t)
\to
\left\{
\ba{rl}
0,&
\theta\in (\phi,\theta_1)\cup(\theta_2,2\pi),\\
 -1,&
\theta\in (\theta_1,\theta_2).
\ea
\right.
\end{equation}
ii) for the $NN$-case,
\begin{equation}\label{p1}
u_d(\rho,\theta,t)
\to
\left\{
\ba{rl}
\ds\fr{2\pi}\Phi-2,&
\theta\in (\phi,\theta_1)\cup(\theta_2,2\pi),\\\\
\ds\fr{2\pi}\Phi-1,&
\theta\in (\theta_1,\theta_2).
\ea
\right.
\end{equation}
iii)  for the $DN$-case,
\begin{equation}\label{p2}
u_d(\rho,\theta,t)
\to
\left\{
\ba{rl}
0,&
\theta\in (\phi,\theta_1),\\
-1,&
\theta\in (\theta_1,\theta_2),\\
-2,&
\theta\in (\theta_2,2\pi).
\ea
\right.
\end{equation}

\end{corollary}
{\bf Proof.} i) First, let us note that
\begin{equation}\label{ui45}
I(\rho,c):={\ds\lim_{t \rightarrow+\infty}}
\int_{-l(t/\rho)}^{l(t/\rho)}\coth (q\beta+ic)d\beta=\frac{i}{2\pi}\ln(-e^{2ic})
\end{equation}
Hence,
\be\la{II}
I(\rho,c)=
\left\{
\ba{rl}
-\ds\frac{c}{\pi}+\ds\frac{1}{2},&
c\in(0,\pi),\\\\
-\ds\frac{c}{\pi}-\ds\frac{1}{2},&
c\in(-\pi,0).
\ea
\right.
\ee
Now (\ref{kjl77}) for the $DD$-problem and (\ref{p1}) for the NN-problem
follow from (\ref{sd1}),  (\ref{ui45}), (\ref{teta12}) and (\ref{9ww}).
The limits (\ref{p2}) follow similarly.\bo

\begin{corollary}
As $t\to\infty$, the total solution $u(y,t)$ of  problem (\ref{sidlu})
\medskip\\
i) Tends to $0$ as $t\to\infty$
for the $DD$ and the $DN$-problems and
\medskip\\
ii) Tends to
 $2\pi/\Phi$ as $t\to\infty$
for the NN-problem.
\end{corollary}
{\bf Proof.}
In the case of the $DD$-problem
\begin{equation}\label{kb5}
u(\rho,\theta,t)\longrightarrow0,~~t\longrightarrow\infty
\end{equation}
since $u_{\rm in}(\rho,\theta,t)\longrightarrow1$ by (\ref{dia}) and
$u_r(\rho,\theta,t)\longrightarrow -1$ for
$\theta\in(\phi,\theta_1)\cup(\theta_2,2\pi)$ by (\ref{1.12}).

 Similarly for $DN$-problem (\ref{kb5}) holds since $u_r\longrightarrow-1$
for $\theta\in(\phi,\theta_1)$ and $u_r\longrightarrow1$ for $\theta\in(\theta_2,2\pi)$ by (\ref{1.13}).

 In the case of the $NN$-problem $u(\rho,\theta,t)\longrightarrow\ds\frac{2\pi}{\Phi}$.
In fact, it follows from (\ref{p1}) and the fact that $u_r(\rho,\theta,t)\longrightarrow1$,
$t\longrightarrow\infty$ for $\theta\in(\phi,\theta_1)\cup(\theta_2,2\pi)$ and
$u_r(\rho,\theta,t)\longrightarrow0$, $t\longrightarrow\infty$ for
$\theta\in(\theta_1,\theta_2)$ by (\ref{1.13}).
\bo


\begin{remark}\label{azz3}
In the case of the $NN$-problem the limit does not vanish and depends on $\Phi$
 in contrast to the case of $DD$ and $DN$- problems.

\end{remark}

\begin{example}\label{3rt1}
Let us consider $F(s)=\delta(s)$. Then ${\hat F}=1$, and
\medskip\\
i) (\ref{udur}) implies that
$u_d$ equals  ${\cal J}_d $ which is given by (\ref{J});
\\
ii) $\hat u_s$
equals
${\cal S}_s=\hat {\cal J}_s$ which
is the solution to (\ref{stdd}).

\end{example}

\section{Sobolev's formula}
\setcounter{equation}{0}

Now we compare our solution to problem (\ref{sidlu})  with the Sobolev
\cite[formula (34)]{Sob34}  solutions.
\begin{lemma}\label{comp}
The solution $u$ to the problem (\ref{sidlu}) given by (\ref{ssi}), (\ref{us}) and (\ref{1.12})
 with the pulse $u_{\rm in}(s)=F(s)=h(s)$
coincides with the Sobolev solution.

\end{lemma}
{\bf Proof.}
We will identify our solution with
the Sobolev one (for the DD-problem) in the form of Petrashen' {\it et al}
\cite[formula (29.3)]{Pet}: for $\tau\in [0,1]$
\begin{equation}\label{U}
\begin{array}{ll}
U(\varphi,\tau)
=\ds\frac{1}{2\pi i}
\Bigg[\ln\ds
\frac{\Big(1-p(\tau)e^{-i\frac{\pi}{\alpha_1}(\varphi-\varphi_{0}+\pi)}\Big)
\Big(1-p(\tau)e^{i\frac{\pi}{\alpha_1}(\varphi-\varphi_{0}-\pi)}\Big)}
{\Big(1-p(\tau)e^{i\frac{\pi}{\alpha_1}(\varphi-\varphi_{0}+\pi)}\Big)
\Big(1-p(\tau)e^{-i\frac{\pi}{\alpha_1}(\varphi-\varphi_{0}-\pi)}\Big)}
\\\\\hspace{3truecm}
-\ln\ds\frac{\Big(1-p(\tau)e^{i\frac{\pi}{\alpha_1}(\varphi+\varphi_{0}-\pi)}\Big)
\Big(1-p(\tau)e^{-i\frac{\pi}{\alpha_1}(\varphi+\varphi_{0}+\pi)}\Big)}
{\Big(1-p(\tau)e^{-i\frac{\pi}{\alpha_1}(\varphi+\varphi_{0}-\pi)}\Big)
\Big(1-p(\tau)e^{i\frac{\pi}{\alpha_1}
(\varphi+\varphi_{0}+\pi)}\Big)}\Bigg],
\end{array}
\end{equation}
 where the branch of $\ln(\cdot)$ is specified by (\ref{b}).
In our notations
\begin{equation}\label{dk}
\begin{array}{l}
\varphi_0=
\alpha -\pi+\al_1,~~\alpha_1=
\Phi,~~
\vp=
\theta  -2\pi+\alpha_1,~~\tau=
\ds\frac{t}{\rho},~~
p(\tau)=b^{-2q}.
\end{array}
\end{equation}
The critical directions $\theta_1$ and $\theta_2$ correspond to
$$
\vp_1=\pi-\vp_0,~~~~~~\vp_2=\pi+2\al_1-\vp_0
$$
Further, in the new variables (\ref{9ww}) reads
\begin{equation}
c_{0} = \ds\frac{\pi}{2\alpha_1}\Big(\varphi-\varphi_{0}+\pi\Big),~~
c_{1} = \ds\frac{\pi}{2\alpha_1}\Big(\varphi+\varphi_0-\pi\Big),~~
c_{2} = \ds\frac{\pi}{2\alpha}\Big(\varphi+\varphi_{0}+\pi-2\alpha_1\Big),~~
 c_{3} = \ds\frac{\pi}{2\alpha}\Big(\varphi-\varphi_0-\pi\Big)
\end{equation}
Substituting these values into (\ref{sd1})
we obtain the diffracted wave in the variables of \ci{Pet}: for   $\tau<1$ and all $\vp\in (0, \alpha_1)$,
\begin{equation}\label{ud1}
\begin{array}{ll}
u_{d}(\varphi,\tau)
=\ds\frac{1}{2\pi i}\Bigg[\ln\ds\frac{1-p(\tau)e^{-i\frac{\pi}{\alpha_1}(\varphi-\varphi_{0}+\pi)}}{-(1-p(\tau)e^{i\frac{\pi}{\alpha_1}(\varphi-\varphi_{0}+\pi)})}
-\ln\ds\frac{1-p(\tau)e^{-i\frac{\pi}{\alpha_1}(\varphi-\varphi_{0}-\pi)}}{-(1-p(\tau)e^{i\frac{\pi}{\alpha_1}(\varphi-\varphi_{0}-\pi)})}
\\\\\hspace{3truecm}+\ln\ds\frac{1-p(\tau)e^{-i\frac{\pi}{\alpha_1}(\varphi+\varphi_{0}-\pi)}}{-(1-p(\tau)e^{i\frac{\pi}{\alpha_1}
(\varphi+\varphi_{0}-\pi)})}-\ln\ds\frac{1-p(\tau)e^{-i\frac{\pi}{\alpha_1}(\varphi+\varphi_{0}+\pi)}}{-(1-p(\tau)e^{i\frac{\pi}
{\alpha_1}(\varphi+\varphi_{0}+\pi)})}\Bigg],
\end{array}
\end{equation}
and $u_d(\varphi,\tau)=0$ for $\tau>1$.
Finally, (\ref{ssi}), (\ref{us}), (\ref{dia})
and (\ref{1.12}) imply that
in the variables of \ci{Pet}
the total solution $u$ to problem (\ref{sidlus}) for the
DD-case is given by
\be\label{u}
\left.
u(\varphi,\tau)=\left\{
\ba{ll}
1+u_d(\varphi,\tau),&\varphi\in (\vp_1,\vp_2)\\
u_d(\varphi,\tau),&\varphi\in (0,\alpha_1)\setminus [\vp_1,\vp_2].
\ea
\right.
\right |
\tau\in[0,1]
\ee

Comparing with (\re{U}), we obtain that
\be\label{uru}
u(\varphi,\tau)= U(\varphi,\tau),~~~\tau\in (0,1), ~~\vp\in (0,\alpha_1)\setminus [\vp_1,\vp_2],
\ee


since $\fr\pi{\al_1}=\fr\pi{\Phi}=q$ by (\re{q}).

For $\varphi\in (\vp_1,\vp_2)$ the coincidence
follows from continuity of the functions $u$ and $U$.

In fact, (\ref{U}), (\ref{ud1}), and (\ref{u}) imply that
for any $(\varphi_0,\tau_0)\in (0,\alpha_1)\times [0,1)$ there exist $\delta (\varphi_0,\tau_0)>0$
and
$k (\varphi_0,\tau_0)\in \Z$
s.t.
$$
U(\varphi,\tau)-u_d(\varphi,\tau)=k((\varphi_0,\tau_0)),~~|(\varphi,\tau)-(\varphi_0,\tau_0)|<\delta.
$$
Hence, $U\equiv u$ by continuity of these functions
since $k=0$ for $\varphi\in (0,\alpha_1)\setminus [\varphi_1,\varphi_2]$
and
$\tau\in [0,1)$
by (\ref{uru}). \bo

\medskip

\begin{remark}\la{rKB}
The coincidence with the Keller-Blank formula requires more cumbersome calculations  and
will be established in other paper.
\end{remark}

\medskip


\section{Appendix}
\setcounter{equation}{0}

{\bf A1} ``Stationary" scattering problem take the forms
\begin{equation}\label{SPNN}
\left\{
	   \begin{array}{rcl}
       (\Delta + \omega^2)\ \hat{u}_s(y,\omega)
            & = & 0,  \hspace{5.7cm} y\in Q   \\ 
       \dfrac{\partial\ }{\partial y_2}\hat{u}_s(y,\omega)
            & = &  -i\omega \hat{F}(\omega) \sin\alpha\  e^{i\omega y_1 \cos\alpha}, \hspace{1.8cm} y\in Q_1 \\ 
       \dfrac{\partial\ }{\partial \mathbf{n_2}}\hat{u}_s(y,\omega)
            & = & i\omega \hat{F}(\omega) \sin(\Phi+\alpha)
                  e^{-i\omega y _2 \frac{\cos(\Phi+\alpha)}{\sin\Phi}},\ \quad y\in Q_2
		\end{array}
\right.
\end{equation}
for the NN-problem and
$$  \left\{
    \begin{array}{lll}
      (-\Delta-\omega^2)\hat{u}_s(y,\omega)=0, \hspace{3.8truecm} y\in Q            \\ \\
      \partial_{y_2}\hat{u}_s(y,\omega)
          = - i \omega g(\omega)\sin \alpha e^{i\omega y_1 \cos \alpha},~\ \ \qquad y\in Q_1    \\ \\
      \hat{u}_s(y,\omega) = -\hat F(\omega) e^{-i\omega y_2 \frac{\cos(\alpha + \Phi)}{\sin \Phi} },
        \hspace{2.6truecm} y \in Q_2
    \end{array}
  \right \vert\quad \omega\in \C^+.
$$
for the DN-problem.

{\bf A2} Malyuzhinetz integral kernels are represented by
\begin{equation}\label{zzic}
H(\beta)=\coth(q(\beta+\frac{i\pi}{2}-i\alpha))\mp\coth(q(\beta-\frac{3i\pi}{2}+i\alpha))
\end{equation}
for $DD$ and $NN$-problems respectively and by
\begin{equation}\label{hih}
H(\beta)=\frac{1}{\sinh[q(\beta+\frac{i\pi}{2}-i\alpha)]}+\frac{1}{\sinh[q(\beta-\frac{3i\pi}{2}+i\alpha)]}
\end{equation}
for $DN$-problem
(see also \cite{MM1}-\cite{eam}) where the DN and NN-problems were considered in details).
Here $q=\frac{\pi}{2\Phi}$.

{\bf A3}.  The densities of the ``stationary" reflected waves  $S_r$
are represented by
\begin{equation}\label{srdn}
\left.
\displaystyle
{ {\cal S}}_{r}(\rho,\theta,\omega)
=
\left\{
\begin{array}{ll}
-e^{i\omega\rho\cos(\theta-\theta_1)}&\phi\leq \theta\leq \theta_1\\
0&\theta_1 <\theta<\theta_2\\
e^{i\omega\rho\cos(\theta-\theta_2)},&\theta_2\leq\theta\leq 2\pi
\end{array}
\right.
\right | \omega\in\overline{\C^+}.
\end{equation}
for the DN-problem and by
\begin{equation}\label{Sr}
\left.
\displaystyle
{ {\cal S}}_{r}(\rho,\theta,\omega)
=
\left\{
\begin{array}{ll}
e^{i\omega\rho\cos(\theta-\theta_1)}&\phi\leq \theta\leq \theta_1\\
0&\theta_1 <\theta<\theta_2\\
e^{i\omega\rho\cos(\theta-\theta_2)},&\theta_2\leq\theta\leq 2\pi
\end{array}
\right.
\right | \omega\in\overline{\C^+}.
\end{equation}
for the NN-problem

{\bf A4}. Inverse Fourier transform.

\begin{lemma}\la{pr/fu/jd}
i) The inverse Fourier transforms of the functions ${\cal S}_d$ and ${\cal S}_r$ given by (\ref{hatjdjr})
are the functions ${\cal J}_d (y,\omega)$ and ${\cal J}_r (y,\omega)$, given
by (\ref{J}).
\end{lemma}
{\bf Proof.} i) We should prove that for $\theta\in\Theta$
\be\la{jod}
{\cal J}_d(\rho,\theta,t)=
F^{-1}_{\omega\to t} {\cal S}_d(\rho,\theta,\omega)= F^{-1}_{\omega\to t} \left [\frac{i}{4\Phi} \int_R e^{i\omega\rho\cosh \beta} Z(\beta+i\theta)d\beta\right].
\ee
First, we note that
$$
\frac{i}{4\Phi}\int_\R e^{i\omega\rho\cosh\beta }Z(\beta+i\theta)d\beta
=\int_0^{\infty} e^{i\omega\rho\cosh\beta }{\cal Z}(\beta+i\theta)d\beta,$$
where
${\cal Z}$ is defined by (\ref{i98}).
The integrals converge by (\ref{ocz}).

\no Changing the variables
$
t:=\rho \cosh\beta,~ d\beta= \frac{dt}{\sqrt{t^2-\rho^2}}
$
we obtain:
\be\la{int1}
\frac{i}{4\Phi}\int_\R e^{i\omega\rho\cosh\beta}Z(\beta+i\theta)d\beta=
F_{t\to\omega}\left [{\cal J} _d(\rho,\theta,t) \right]
\ee
where
${\cal J} _d$ is given by (\ref{J}).
Hence, (\ref{jod}) follows.

\no The representation (\ref{Sr}) for $ {\cal J}_r$ follows from (\ref{J}) directly.
\bo

The representations for ${\cal J}_r$ for other cases:
\be\la{jotardn}
{\cal J}_r(\rho,\theta,t):=
\left\{
\ba{ll}
\displaystyle
\delta(t-\rho\cos(\theta-\theta_1)),&\theta\in (\phi,\theta_1)\\
0, & \theta\in (\theta_1,\theta_2)\\
-\delta(t-\rho\cos(\theta-\theta_2)),&\theta\in (\theta_2,2\pi)\\
\ea
\right| t\geq 0,~~
{\cal J}_r(\rho,\theta,t)=0,~  t<0.
\\
\ee
for the DN-problem and
\be\la{jotarnn}
{\cal J}_r(\rho,\theta,t):=
\left\{
\ba{ll}
\displaystyle
-\delta(t-\rho\cos(\theta-\theta_1)),&\theta\in (\phi,\theta_1)\\
0, & \theta\in (\theta_1,\theta_2)\\
-\delta(t-\rho\cos(\theta-\theta_2)),&\theta\in (\theta_2,2\pi)\\
\ea
\right| t\geq 0,~~
{\cal J}_r(\rho,\theta,t)=0,~  t<0.
\\
\ee

For the DN-problem

{\bf A5}. Expressions for the reflected waves

 \begin{equation} \label{1.12}
u_r(y,t):=\left\{
\begin{array}{ll}
\mp F(t-n_1\cdot y),&\phi   \leq \theta\leq  \theta_1\\
0&\theta_1< \theta<\theta_2,\\
\mp F(t-n_2\cdot y),&\theta_2 \leq \theta \leq 2\pi,\\
\end{array}
\right.
 \end{equation}
for DD and NN-problems respectively and
\begin{equation} \label{1.13}
u_r(y,t):=\left\{
\begin{array}{ll}
F(t-n_1\cdot y),&\phi   \leq \theta\leq  \theta_1\\
0&\theta_1< \theta<\theta_2,\\
-F(t-n_2\cdot y),&\theta_2 \leq \theta \leq 2\pi,\\
\end{array}
\right.
 \end{equation}
for the DN-problem.
\vspace {1cm}

{\bf A6}. Jumps of the diffracted wave in the cases of NN and DN problems.

For the DN-problem the function $Z$ from (\ref{n2} takes the form:
\begin{equation}\label{zdn}
\begin{array}{ll}
Z(\beta+i\theta)=-\ds\frac{1}{\sinh\Big(q\beta+ic_0\Big)}-\ds\frac{1}{\sinh\Big(q\beta+ic_1\Big)}-\ds\frac{1}{\sinh\Big(q\beta+ic_2\Big)}+\ds\frac{1}{\sinh\Big(q\beta+ic_3\Big)}
\end{array}
\end{equation}

The jumps of the diffracted wave $u_d$ on the critical rays for the NN and DN-problems
are given by
\be\label{jkk}
[u_d]_k(\rho,t)=F(t-\rho),
~~~\rho>0,~k=1,2
\ee

(cf. with (\ref{jk})).


\medskip








\begin{thebibliography}{99}


\bibitem{kmr} A.I. Komech, A.E.Merzon, Relation between Cauchy data
for the scattering by a wedge, {\em Russian Journal of Mathematical
Physics} {\bf 14} (2007) no. 3, 279-303.




\bibitem{kmm}  Komech AI , Mauser NJ, Merzon AE.
On Sommerfeld representation and uniqueness in scattering by wedges.
{\em Mathematical Methods in the Applied Sciences}  {\bf
28} (
)(2005), 147-183.



\bibitem{km} A.I.Komech, A.E.Merzon.
Limiting Amplitude Principle in the Scattering by Wedges. {\it
Mathematical Methods in the Appliied Sciences} {\bf 29} (2006) , 1147-1185





\bibitem{MM1}J.E.de la Paz Mendez, A.Merzon, Scattering of a plane wave by hard-soft wedges.
{\ em Recent Progress in Operator Theory and Its Applications}. Series: Operator Theory: Advances and Applications,
(2012), {\bf  220}, 207-227.


\bibitem{md3} A. Merzon. Well-posedness  of the problem of
nonstationary diffraction of Sommerfeld. {\it Proceeding of the
International Seminar "Day on Diffraction-2003"}. University of
St.Petersburg (2003) 151-162.



\bibitem{mm} J.E de la Paz Mendez, A.E. Merzon, DN-Scattering of a plane wave by wedges.
{\em Mathematical Methods in the Applied Sciences}, {\bf 34}, No. 15, (2011), 1843-1872 (http://onlinelibrary.wiley.com/doi/10.1002/mma.1484/abstract).

\bibitem{yet7} Anatoli Merzon, Jos\'e Eligio De la Paz M. {\it DN-Problema de dispersi\'on de una onda plana sobre una cu\~na. Principio de Amplitud L\'imite.}
Editorial Acad\'emica Espa\~nola (2012-10-03)

\bibitem{eam} A.Esquivel, A.E.Merzon. NN-problem (in preparation)


\bibitem{Sob34} S.L. Sobolev,
Theory of diffraction of plane waves,
Proceedings of Seismological Institute, no. 41,
Russian Academy of Science, Leningrad, 1934.




\bibitem{Sob}  S.L. Sobolev,
General theory of diffraction of waves on Riemann surfaces,
{\em Tr. Fiz.-Mat. Inst. Steklova} {\bf 9} (1935), 39-105.
[Russian] (English translation:
S.L. Sobolev, General theory of diffraction of waves on Riemann surfaces,
p. 201-262 in: Selected Works of S.L. Sobolev, Vol. I,
Springer, New York,  2006.)



 \bibitem{Sob1} S.L. Sobolev,
Some questions in the theory of propagations of oscillations,
Chap XII, in: {\it Differential
  anf Integral Equations of Mathematical Physics},
F.Frank and P. Mizes (eds), Leningrad-Moscow (1937) pp 468-617.[Russian]




\bibitem{KB} Keller J, Blank A. Diffraction
 and reflection of pulses by wedges and corners.
 {\em Communications on Pure and Applied Mathematics} 1951; {\bf
4}(1):75-95.



\bibitem{Ka} Kay I. The diffraction of an arbitrary pulse
 by a wedge. {\em Communications on Pure and Applied Mathematics} 1953; {\bf 6}:521-546.






\bibitem{3}  Komech AI. Elliptic boundary value problems on manifolds
 with piecewise smooth boundary.
{\em Math. USSR Sbornik}. 1973; {\bf 21}(1): 91-135.





\bibitem{5}  Komech AI, Merzon AE.
 General boundary value problems in
 region with corners. {\em Operator Theory: Adv.
 Appl.} 1992; {\bf 57}: 171-183.






\bibitem{KMZ}  Komech A, Merzon A, Zhevandrov P. A method of
 complex characteristics for elliptic problems in angles and its
 applications. {\em American Mathematical Society Translation} 2002; {\bf
 206}(2):125-159.




\bibitem{cmkz}
A. Choque,  Yu. Karlovich, A. Merzon and P. Zhevandrov. On the convergence of the amplitude of the diffracted nonstationary wave in scattering by wedges. Russian Journal of Mathematical Physics, 2012, V.19, N 3, pp.373-384.











 \bibitem {Pet} Petrashen' CI, Nikolaev VG, Kouzov
 DP.
On the series method  in the theory of diffraction of waves by
polygonal regions. {\em Nauchnie ZapiFi LGU.} 1958;
  {\bf 246}(5): 5-70 (in Russian).












 \bibitem{Fe} Felsen LB, Marcuvitz N.
 {\em Radiation and Scattering of Waves.} Oxford Univ Pr: Oxford,
  1996.



\bibitem {Ro}  Rottbrand K. Time-dependent plane wave diffraction
 by a half-plane: explicit solution for Rawlins' mixed initial
 boundary value problem. {\em Z.Angew. Math. Mech.} 1998;
 {\bf 78}(5): 321-335.


\bibitem{MPR}  Meister E, Passow A, Rottbrand K.
New results on wave diffraction by canonical obstacles. {\em
Operator Theory: Adv. Appl.} 1999; {\bf 110}: 235-256.


\bibitem {Rot}  Rottbrand K. Exact solution for time-dependent
  diffraction of plane waves by semi-infinite soft/hard wedges
  and half-planes. Preprint 1984 Technical University  Darmstadt.
  1998

\bibitem{bor} Borovikov V.A. {\em Diffracion at Poligons and Polyhedrons.} Nauka, Moscow.(1966)



\bibitem{41} Komech A.I. Linear partial differential equations with
constant coefficients. In Egorov YuE, Komech AI, Shubin MA. {\em
Elements of the Modern Theory of Partial Differential Equations}.
Springer: Berlin, 1999: 127-260.


\bibitem{GS} Gel'fand I.M., Shilov G. E. {\em Generalized functions. Vol. 2. Spaces of fundamental and generalized functions}, Boston, MA, (1968)





\bibitem{111}  Merzon AE. On Ursell's problem.
{\em Proceedings of the Third International Conference on
 Mathematical and Numerical Aspects of Wave Propagation}.
 SIAM -- INRIA, edited by Gary Cohen. 1995; 613-623.

\bibitem{Laur5}  Komech A.I.,  Merzon A.E., Zhevandrov P.N.,
 On completeness of Ursell's trapping modes.
{\em  Russian Journal of
 Mathematical Physics}. 1996; {\bf 4}(4): 457-485.


\bibitem{fed}
Fedoryuk MV. {\em Asymptotics:  Integrals and Series.} Nauka:
Moscow, 1987.



\bibitem{Vlad}


Vladimirov, V. S. (1979), Generalized functions in mathematical physics (in English), Moscow: Mir Publishers, p. 362, ISBN 0-8285-0001-0, MR 0564116, Zbl 0515.46034. A textbook on the theory of generalized functions and their applications to mathematical physics and several complex variables.

\bibitem{Vlad1}

Vladimirov, V.S. (1983), Equations of mathematical physics (in English) (2nd ed.), Moscow: Mir Publishers, p. 464, MR 0764399, Zbl 0207.09101 (Zentralblatt review of the first English edition).




V.S.Vladimirov



A. Choque,  Yu. Karlovich, A. Merzon and P. Zhevandrov. On the convergence of the amplitude of the diffracted nonstationary wave in scattering by wedges. Russian Journal of Mathematical Physics, 2012, V.19, N 3, pp.373-384.



\bibitem{WW}E.L. Whittaker, G.N. Watson,
A Course of Modern Analysis, Macmillan, New York, 1948.


 \end{thebibliography}
\end{document}